\documentclass[conference]{IEEEtran}
\usepackage{amsmath,amsfonts}
\usepackage{algorithmic}
\usepackage{hyperref}
\usepackage{algorithm}
\usepackage{array}
\usepackage{color,soul}
\usepackage{multirow}
\usepackage{tabularx}
\usepackage{threeparttable}
\usepackage{amssymb}
\usepackage{pifont}
\usepackage{textcomp}
\usepackage{stfloats}
\usepackage{url}
\usepackage{booktabs}
\usepackage{hhline}
\usepackage{verbatim}
\usepackage{graphicx}
\usepackage{cite}
\usepackage{diagbox}
\usepackage{lettrine}  

\usepackage{orcidlink}
\begin{document}

\title{Energy Efficient Exact and Approximate Systolic Array Architecture for Matrix Multiplication}
\author{
\IEEEauthorblockN{
Pragun Jaswal\textsuperscript{*}~\orcidlink{0009-0005-9580-3184},
L.~Hemanth Krishna\textsuperscript{†}~\orcidlink{0000-0002-7737-6685},
B.~Srinivasu\textsuperscript{‡}~\orcidlink{0000-0003-0974-8245}
}
\IEEEauthorblockA{
Indian Institute of Technology Mandi, Mandi--175005, India\\
\textsuperscript{*}thepragun@gmail.com,
\textsuperscript{†}hemanth.krishna412@gmail.com,
\textsuperscript{‡}srinivasu@iitmandi.ac.in
}
}

\maketitle
\begin {abstract}
Deep Neural Networks (DNNs) require highly efficient matrix multiplication engines for complex computations. 
This paper presents a Systolic Array (SA) architecture incorporating novel exact and approximate Processing Elements (PEs), designed using 
energy-efficient partial product and nand based partial product cells, termed as PPC and NPPC, respectively. 
The proposed 8-bit exact and approximate PE designs are employed in a 8x8 SA, which achieves a energy savings of 16\% and 68\%, respectively, compared to the existing design.
To demonstrate their effectiveness, the proposed signed PEs are integrated into a 
SA for Discrete Cosine Transform (DCT) computation, 
achieving high output quality with a PSNR of 45.97\,dB. 
Furthermore, for edge detection applications using both kernel-based and CNN-based approaches, the proposed approximate signed PEs achieve PSNR values of 30.45 dB and 75.98 dB, respectively. These results highlight the potential of the proposed design to deliver significant energy efficiency 
while maintaining competitive output quality, 
making it well-suited for error-resilient image and vision processing applications.
\end{abstract}


\newtheorem{remark}{Remark}
\newtheorem{Theorem}{Theorem}
\newtheorem{Proposition}{Proposition} 
\newtheorem{Proof}{Proof} 

\section{Introduction}

\lettrine[lines=2, lhang=0.1, loversize=0]{\textbf{A}}{rtificial} Intelligence (AI) and Machine Learning (ML) have become integral to a wide range of applications, 
including computer vision, natural language processing, 
speech recognition, and autonomous systems~\cite{lecun2015deep, goodfellow2016deep}. 
At the core of these workloads are deep neural networks (DNNs), which rely 
extensively on large-scale matrix multiplications to perform 
computations across multiple layers of interconnected neurons. 
Both the training and inference phases of DNNs involve billions 
of Multiply–ACcumulate (MAC) operations, 
making matrix multiplication one of the most 
computationally and energy-intensive kernels in modern AI accelerators~\cite{Mittal2022}.

Systolic arrays (SAs) have emerged as an effective 
hardware architecture for 
accelerating matrix multiplications~\cite{Ahmed2025}. 
Recent advances in domain-specific hardware, such as 
Google's Tensor Processing Unit (TPU), 
have demonstrated the ability of 
systolic arrays to deliver 
high throughput for DNN workloads. 
However, conventional systolic arrays 
typically employ exact arithmetic units, 
which incur significant energy 
consumption and large silicon area overheads. 
These limitations hinder their deployment in 
resource-constrained platforms, such as edge devices 
and Internet of Things (IoT) systems, where energy efficiency 
is of paramount importance.
To mitigate these challenges, approximate computing has been widely explored as a design paradigm that trades exactness for improvements in energy efficiency and hardware complexity.

In this paper, we propose both exact and approximate processing element that supports both signed and unsigned operations for systolic array-based matrix multiplication. The proposed PE significantly reduces hardware complexity and energy consumption while maintaining acceptable image quality. This makes it well-suited for error-tolerant applications such as image processing and machine learning, where efficiency is prioritized over exact precision. The major contributions of this paper are summarized as follows:
\begin{itemize}
    \item Proposed an energy-efficient approximate systolic array architecture for matrix multiplication, introducing novel exact and approximate Processing Elements (PEs). 
    \item Proposed approximate Partial Product Cell (PPC) and Nand based Partial Product Cell (NPPC), achieve 46.8\% and 34.4\% of energy savings respectively as compared to the best existing design \cite{Haroon2021}.
    \item Proposed exact signed 8-bit PE achieves 24.37\% of energy saving over the design \cite{Lombardi2015}, while the proposed approximate signed 8-bit PE achieves 22.51\% of energy saving over the design \cite{Haroon2021}.
    \item Validation of the proposed architecture on three applications: DCT for image compression, kernel-based edge detection and CNN-based edge detection, results have been found to achieve a PSNR of 45\,dB for DCT, 30.45\,dB for kernel-based edge detection and 75.98\,dB for CNN-based edge detection.
\end{itemize}
The rest of the paper is organized as follows. Section \ref{sec:systolic} explains the matrix multiplication using systolic array architecture. Section \ref{sec:PE} presents the proposed processing elements for the matrix multiplication. Sections \ref{Sec:results} and \ref{sec:Application} present
the results and applications, respectively. The conclusions are presented in Section \ref{sec:conclusion}.

\section{Systolic Array Based Matrix Multiplication Architecture}
\label{sec:systolic}

Matrix multiplication plays a major role in accelerating the applications such as machine learning, signal processing, and neural networks. Several systolic array architectures for matrix multiplication are found in literature \cite{Kung1982WhySA, Ortega2024, varman1986synthesis}.  Recently, with applications in neural networks with larger data, GEMM-based matrix multiplication using a systolic array is highly appreciated 
for machine learning applications \cite{Inayat2022}, \cite{Ortega2024}. 

\begin{figure}[H]
	\centerline{\includegraphics[scale=0.45]{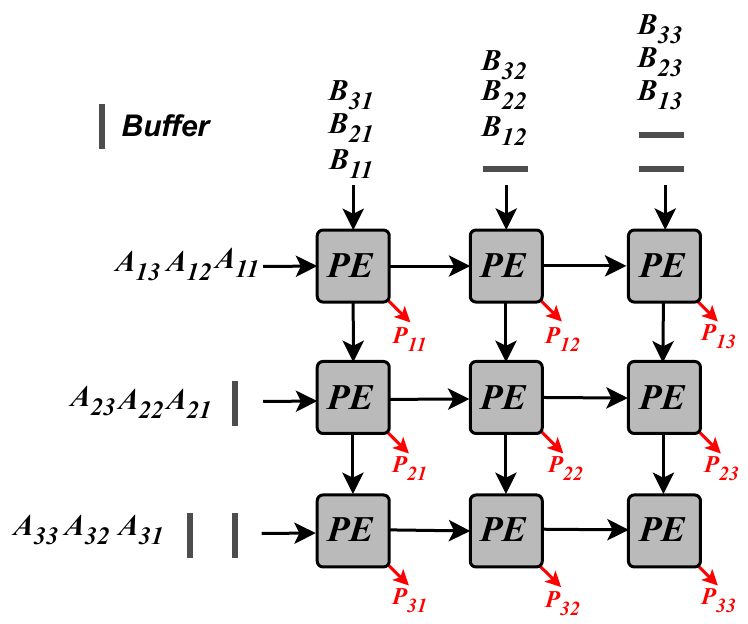}}
        \caption{Architecture of $3 \times 3$ Systolic Array for Matrix Multiplication\cite{Kung1982WhySA}.}
        \label{fig:mat_mul}
\end{figure}

The latency of the conventional systolic array architecture, illustrated in Fig. \ref{fig:mat_mul}, is given by $3N - 2$ clock cycles \cite{Latency}. 

\begin{figure}[H]
    \centering
    \includegraphics[width=0.8\linewidth]{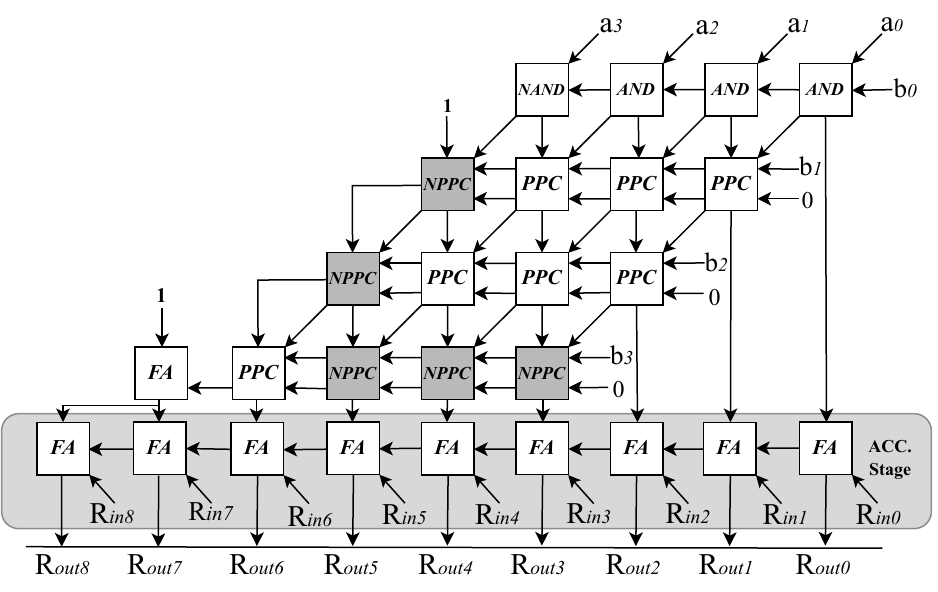}
    \caption{Existing Design of Exact 4-bit Signed PE \cite{Lombardi2015}.}
    \label{fig:exact_pe}
\end{figure}

\begin{figure}[H]
    \centering
    \includegraphics[width=1\linewidth]{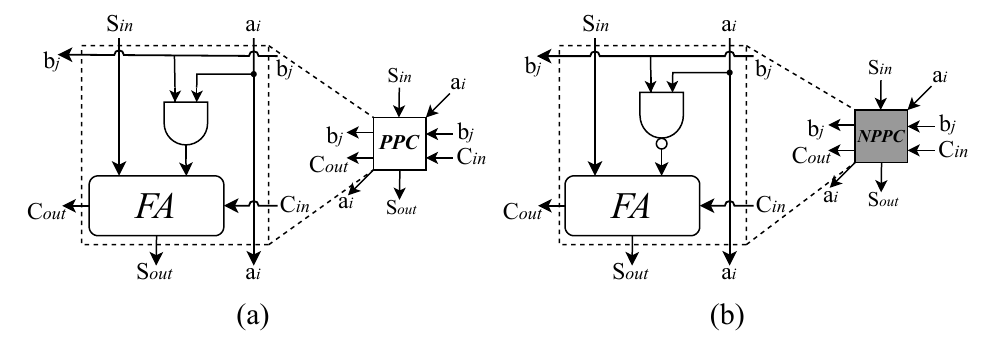}
    \vspace{-7mm}
    \caption{Conventional approach of Exact (a)  PPC and (b) NPPC. }
    \label{fig:conv_ppc}
\end{figure}

In a conventional systolic array as shown in Fig. \ref{fig:mat_mul}, the PE is the 
basic block for executing MAC operations \cite{varman1986synthesis}. 
Since a large number of PEs operate concurrently, it has a critical impact on the overall performance, energy efficiency, and area utilization. 
In the conventional design, a PE typically consists of a single multiplier followed by an adder to perform 
the MAC operation \cite{varman1986synthesis}. 
Alternatively, some studies optimize the PE by employing PPC 
and NPPC \cite{Lombardi2015}, to generate positive partial products and  
negative partial products, respectively, for enabling efficient signed multiplication as presented in 
Fig. \ref{fig:exact_pe}. The partial products are then accumulated using full adders (FA) in the accumulation stage.
Fig.~\ref{fig:conv_ppc} illustrates the exact PPC and NPPC blocks used to perform the bit-wise MAC 
operations.

The next section presents the proposed processing elements for performing the matrix multiplication. 

\section{Hardware Implementation of Proposed Processing Element (PE)}
\label{sec:PE}

Firstly, we present the exact PE architecture using the proposed optimized PPC and NPPC cells then approximate PE architecture using approximate PPC and NPPC cells.

\subsection{Proposed Exact Processing Element}
\begin{figure}[H]
    \centering
    \includegraphics[width=1\linewidth]{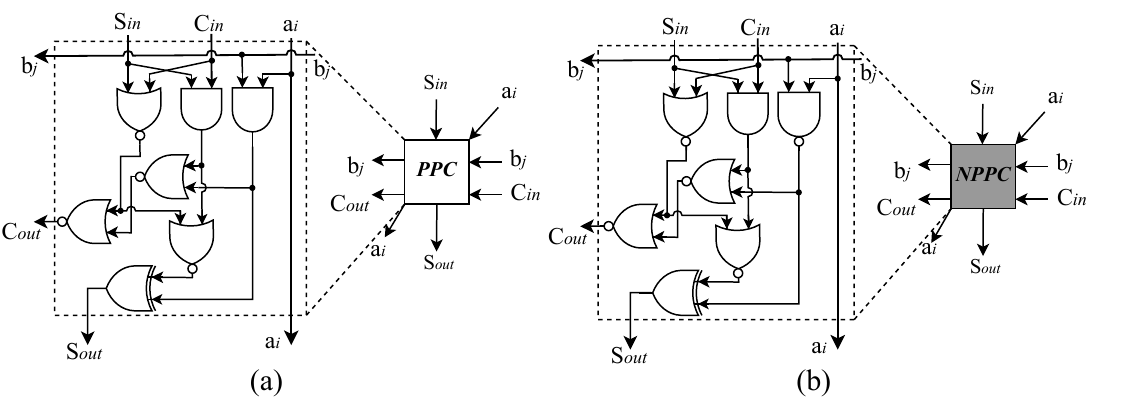}
    \vspace{-4mm}
    \caption{Proposed  exact designs of (a) PPC and (b) NPPC }
    \vspace{-3mm}
    \label{fig:prop_ppc}
\end{figure}

\begin{figure}[H]
    \centering
    \includegraphics[width=1\linewidth]{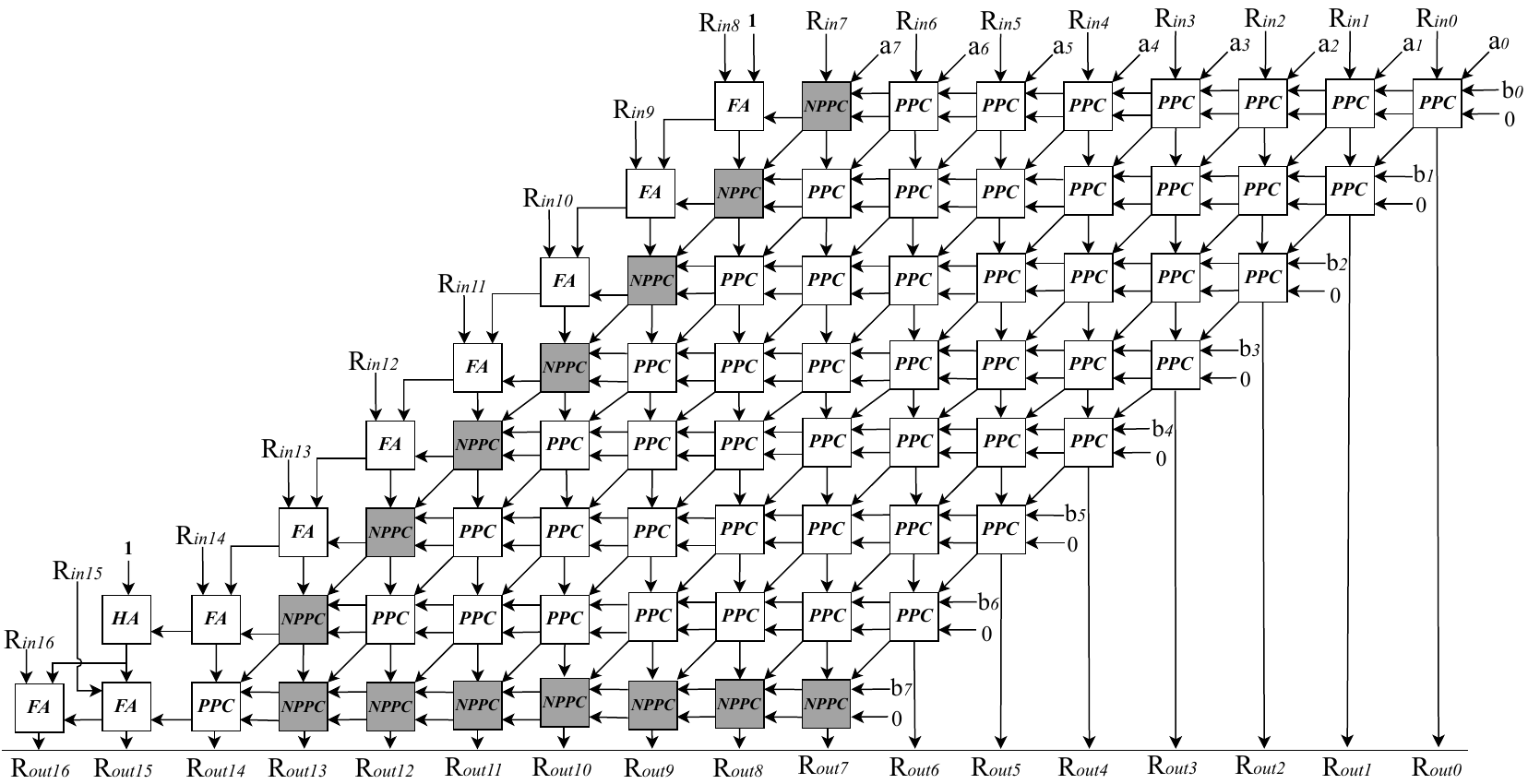}
    \vspace{-4mm}
    \caption{Proposed exact 8-bit signed PE Design.}
    \vspace{-4mm}
    \label{fig:exact_prp_s8}
\end{figure}

\begin{figure*}
    \centering
    \includegraphics[width=1\linewidth]{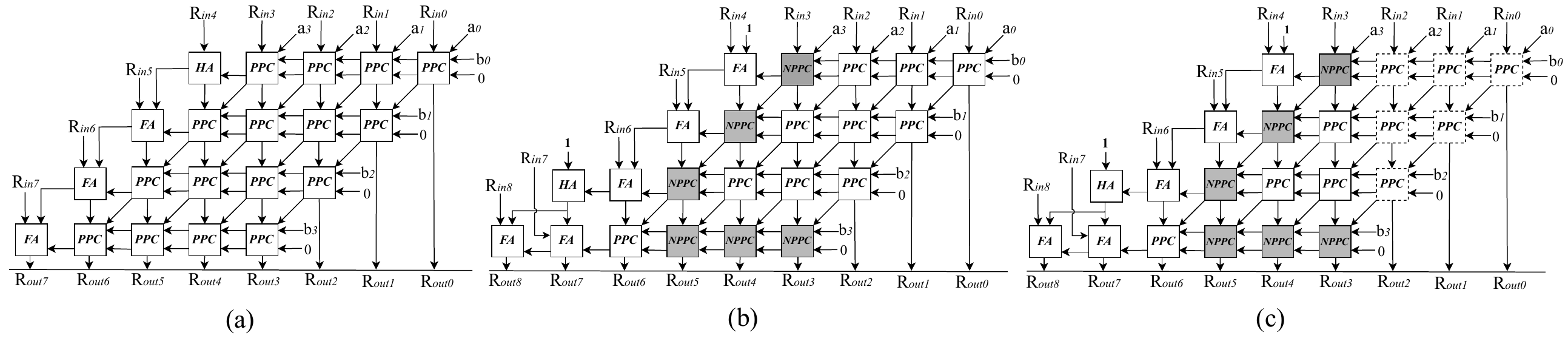}
    \vspace{-5mm}
    \caption{Proposed 4-bit PE designs (a) exact unsigned, (b) exact signed, and (c) approximate signed $(k=N-1)$ .}
    \vspace{-5mm}
    \label{fig:main_design}
\end{figure*}

The proposed PE architecture eliminates the conventional separation between 
multiplication and accumulation by integrating them into a unified processing element to perform 
$a \times b + c$. 
This fused approach enables simultaneous reduction of both partial products and 
the accumulated sum, improving efficiency and reducing delay.
The proposed exact PPC and NPPC for the PE are shown in Fig. \ref{fig:prop_ppc} (a) and (b), respectively.
Fig. \ref{fig:exact_prp_s8} and \ref{fig:main_design} (b) illustrates the exact 8-bit and 4-bit signed PE architecture, respectively. The NPPC cells play a crucial role in handling signed operations.



Fig. \ref{fig:main_design} (a) illustrates the exact 4-bit unsigned PE. 
In this architecture, PPC play a crucial role in performing the MAC operation. 
For the 8-bit signed PE, a total of 14 NPPC cells and 50 PPC cells are required, whereas the 
existing work in \cite{Lombardi2015} requires 15 number of additional full adders as well. 
In general, for an $N$-bit PE, the implementation requires $N^2 - 2N - 2$ PPC cells and $2N - 2$ NPPC cells.

\subsection{Proposed Approximate Processing Element}
\begin{figure}[H]
    \centering
    \includegraphics[width=1\linewidth]{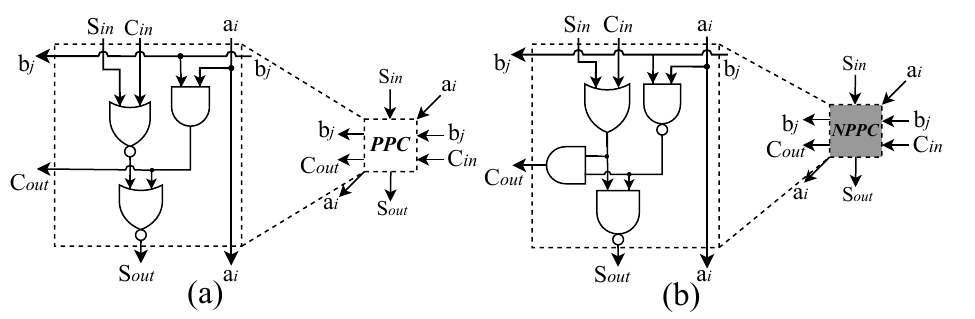}
    \vspace{-5mm}
    \caption{Proposed Approximate designs of  (a)  PPC and (b) NPPC. }
    \vspace{-4mm}
    \label{fig:prop_appc}
\end{figure}

In the proposed design, the PE is constructed using approximate PPC and NPPC blocks. 
Fig. \ref{fig:main_design} (c) illustrates the architecture of the proposed approximate 4-bit signed PE, where the dotted blocks represent the approximate PPC and NPPC units for approximate factor $k = N-1$.
The proposed approximate PPC and NPPC are presented in 
Fig. \ref{fig:prop_appc}, and their truth tables are provided in 
Table \ref{tab:ppc_truth}.  
    

The Boolean expression for the \textit{Sum} output of the proposed approximate PPC cell is 
$S_{\text{out}} = \overline{\overline{(S_{\text{in}} + C_{\text{in}})} + (a_i \cdot b_i)}$,
while the corresponding carry output is 
$C_{\text{out}} =  a_i \cdot b_i$.
For the NPPC block, the \textit{Sum} output is expressed as  
$S_{\text{out}} = \overline{(S_{\text{in}} + C_{\text{in})} \cdot \overline{(a_i \cdot b_i)}}$,
and the carry output is given by
$C_{\text{out}} = (S_{\text{in}} + C_{\text{in})} \cdot \overline{(a_i \cdot b_i)}$.
The proposed approximate PPC introduces five erroneous outputs, out 
of a total of 16 cases, resulting in an error rate of $5/16$.  
The error occurs for the following combinations of inputs,  $(a_i, b_i, S_{\text{in}}, C_{\text{in}})$ are 
$(0,0,1,1)$, $(0,1,1,1)$, $(1,0,1,1)$, $(1,1,0,0)$, and $(1,1,1,1)$.  
The corresponding error probabilities ($P_E$) for these cases are $9/256$, $3/256$, $3/256$, $9/256$, and $1/256$, respectively.  
Therefore, the total error probability of the proposed PPC and NPPC blocks is $25/256$.

\begin{table}
\centering
\scriptsize
\caption{Truth Table for Partial Product Cell (PPC) and Negative Partial Product Cell (NPPC)}
\setlength{\tabcolsep}{3.5pt}  
\renewcommand{\arraystretch}{1.2}
\label{tab:ppc_truth}
\begin{tabular}{|c c c c||c c|c c|c||c c|c c|c|}
\hline
\multicolumn{4}{|c||}{\multirow{2}{*}{\textbf{Inputs}}} 
& \multicolumn{5}{c||}{\textbf{PPC: $a \cdot b + C_{in} + S_{in}$}} 
& \multicolumn{5}{c|}{\textbf{NPPC: $\overline{a \cdot b} + C_{in} + S_{in}$}} \\
\cline{5-14}
& & & & \multicolumn{2}{c|}{\textbf{Exact}} & \multicolumn{3}{c||}{\textbf{Approx}} 
& \multicolumn{2}{c|}{\textbf{Exact}} & \multicolumn{3}{c|}{\textbf{Approx}} \\
\hline
$a_i$ & $b_i$ & $C_{in}$ & $S_{in}$ & $C$ & $S$ & $C$ & $S$ & \textbf{ED} 
& $C$ & $S$ & $C$ & $S$ & \textbf{ED} \\
\hline
0 & 0 & 0 & 0 & 0 & 0 & 0 & 0 & 0 & 0 & 1 & 0 & 1 & 0 \\
0 & 0 & 0 & 1 & 0 & 1 & 0 & 1 & 0 & 1 & 0 & 1 & 0 & 0 \\
0 & 0 & 1 & 0 & 0 & 1 & 0 & 1 & 0 & 1 & 0 & 1 & 0 & 0 \\
0 & 0 & 1 & 1 & 1 & 0 & 0 & 1 & \textcolor{red}{-1} & 1 & 1 & 1 & 0 & \textcolor{red}{-1} \\
\hline
0 & 1 & 0 & 0 & 0 & 0 & 0 & 0 & 0 & 0 & 1 & 0 & 1 & 0 \\
0 & 1 & 0 & 1 & 0 & 1 & 0 & 1 & 0 & 1 & 0 & 1 & 0 & 0 \\
0 & 1 & 1 & 0 & 0 & 1 & 0 & 1 & 0 & 1 & 0 & 1 & 0 & 0 \\
0 & 1 & 1 & 1 & 1 & 0 & 0 & 1 & \textcolor{red}{-1} & 1 & 1 & 1 & 0 & \textcolor{red}{-1} \\
\hline
1 & 0 & 0 & 0 & 0 & 0 & 0 & 0 & 0 & 0 & 1 & 0 & 1 & 0 \\
1 & 0 & 0 & 1 & 0 & 1 & 0 & 1 & 0 & 1 & 0 & 1 & 0 & 0 \\
1 & 0 & 1 & 0 & 0 & 1 & 0 & 1 & 0 & 1 & 0 & 1 & 0 & 0 \\
1 & 0 & 1 & 1 & 1 & 0 & 0 & 1 & \textcolor{red}{-1} & 1 & 1 & 1 & 0 & \textcolor{red}{-1} \\
\hline
1 & 1 & 0 & 0 & 0 & 1 & 1 & 0 & \textcolor{red}{+1} & 0 & 0 & 0 & 1 & \textcolor{red}{+1} \\
1 & 1 & 0 & 1 & 1 & 0 & 1 & 0 & 0 & 0 & 1 & 0 & 1 & 0 \\
1 & 1 & 1 & 0 & 1 & 0 & 1 & 0 & 0 & 0 & 1 & 0 & 1 & 0 \\
1 & 1 & 1 & 1 & 1 & 1 & 1 & 0 & \textcolor{red}{-1} & 1 & 0 & 0 & 1 & \textcolor{red}{-1} \\
\hline
\end{tabular}
\end{table}

\section{Results and Discussion}
\label{Sec:results}
This section presents hardware evaluation of the proposed and existing designs in terms of area, power, and delay 
using the Cadence Genus synthesis solution. In addition, error metrics are analyzed through Python-based simulations.

\subsection{Hardware Evaluation}
The existing and proposed designs are designed using Verilog HDL, and then the designs are 
synthesized using the Cadence Genus Synthesis solution using a 90-nm UMC technology.
The synthesized results were then evaluated based on three primary hardware metrics: critical path delay, power consumption, and area, as summarized in Table~\ref{tab:ppc} for the proposed PPC and NPPC. The analysis shows an energy improvement of 6.4\% for the exact design \cite{Lombardi2015}, and a significant 46.8\% improvement for the proposed approximate PPC design compared to the best existing design \cite{Waris2019}.

\begin{table}
\centering
\scriptsize
\setlength{\tabcolsep}{2.8pt} 
\renewcommand{\arraystretch}{1.1}  
\caption{Hardware Analysis of Proposed and Existing PPCs and NPPCs }
\label{tab:ppc}
\begin{tabular}{ccccccccc}
\toprule
\multirow{4}{*}{\large{Design}} & \multicolumn{4}{c}{\textbf{PPC}} & \multicolumn{4}{c}{\textbf{NPPC}} \\
\cmidrule(lr){2-5} \cmidrule(lr){6-9}
  & Area & Power & Delay & PDP & Area & Power & Delay & PDP \\
  & ($\mu$m$^2$) & ($\mu$W) & (ps) & (aJ) & ($\mu$m$^2$) & ($\mu$W) & (ps) & (aJ) \\
\midrule
Exact  \cite{Lombardi2015}     & 25.81 & 1.03 & 262 & 269.86 & 24.92 & 0.99 & 238 & 235.62 \\
Prop Ext                       & 24.98 & 0.99 & 255 & 252.45 & 23.47 & 0.99 & 216 & 213.84 \\
Design \cite{Lombardi2015}     & 13.32 & 0.64 & 187 & 119.04 & 12.54 & 0.61 & 156 & 95.16 \\
Design \cite{Haroon2021}       & 14.13 & 0.58 & 157 & 91.06  & 13.22 & 0.60 & 148 & 88.80 \\
Prop Apx                       & 10.19 & 0.44 & 110 & 48.4   & 9.40  & 0.37 & 147 & 54.39 \\
\bottomrule
\end{tabular}
\end{table}

Table~\ref{tab:pe} presents the analysis of the proposed and existing PEs for both unsigned and signed designs. The results indicate that the proposed exact PEs achieve superior efficiency compared to existing exact designs, with notable reductions in area, power, and delay, leading to overall improvements of up to 16\%.
Table~\ref{tab:pe} also includes PE designs without PPC and NPPC blocks \cite{Gemini2021,Shaf2023}. A significant improvement in the Power Area Delay Product (PADP) of 65.45\% is observed when compared to the design in \cite{Gemini2021}, highlighting that PPC and NPPC based PE designs are highly efficient in terms of hardware metrics.
Furthermore, the proposed approximate PE achieve up to 23\% improvement in the PADP compared to the best existing approximate design \cite{Haroon2021}.

\begin{table}
\centering
\scriptsize
\setlength{\tabcolsep}{2.0pt}
\renewcommand{\arraystretch}{1.1}
\caption{Hardware Analysis of Proposed and Existing Processing Elements }
\label{tab:pe}
\begin{threeparttable}
\begin{tabular}{cccccccccc}
\toprule
\multirow{3}{*}{\textbf{PE}} & \multirow{2}{*}{\textbf{bit}} &\multicolumn{4}{c}{\textbf{Unsigned}} & \multicolumn{4}{c}{\textbf{Signed}} \\
\cmidrule(lr){3-6} \cmidrule(lr){7-10}
\multirow{2}{*}{ \textbf{Design}} & \multirow{1}{*}{\textbf{Width}}&Area & Power & Delay & PADP$^{\delta}$ & Area & Power & Delay & PADP$^{\delta}$\\
  & $(N)$ &($\mu$m$^2$) & ($\mu$W) & (ns) & ($\times 10^3$) & ($\mu$m$^2$) & ($\mu$W) & (ns) & ($\times 10^3$) \\
  
\midrule
\multicolumn{10}{c}{\textbf{Exact PPCs and NPPCs based Designs}} \\
\midrule

Design
  &{4} 
    & 435.9 & 29.4 & 1.87 & 23.96 & 446.5 & 29.7 & 1.65 & 21.82 \\
\cite{Lombardi2015}  
& {8} 
    & 1718.5 & 181.3 & 3.92 & 1222.57 & 1708 & 183.4 & 3.71 & 1162.39 \\

\midrule

Design
  &{4} 
    & 432.8 & 30.4 & 1.76 & 23.13 & 445.3 & 31.7 & 1.55 & 21.88 \\
\cite{Haroon2021}  & {8} & 1730.6 & 185.3 & 3.67 & 1175.71 & 1716 & 190.3 & 3.22 & 1050.21 \\
\midrule

\multirow{2}{*}{Proposed} 
  &{4} 
    & 411 & 26.6 & 1.73 & 18.91 & 419 & 26.8 & 1.52 & 17.06 \\
  & {8} 
    & 1659.2 & 180.7 & 3.65 & 1094.33 & 1620.3 & 170.6 & 3.18 & 879.02\\

\midrule
\multicolumn{10}{c}{\textbf{Conventional Approach Exact MAC Designs}} \\
\midrule

HA-FSA\cite{Inayat2022}$^{*}$  & 8 & - & - & - & - & 2012 & 465 & 2.3 & 1662.1 \\
\midrule
Gemmini\cite{Gemini2021}$^{*}$  & 8 & - & - & - & - & 1968 & 344 & 2.9 & 1763.7 \\
\midrule
\multicolumn{10}{c}{\textbf{Approximate PPCs and NPPCs based Designs ($k=N-1$)}} \\
\midrule
\multirow{2}{*}{Design \cite{Lombardi2015}} 
  &{4} 
    & 416.30 & 24.1 & 1.56 & 15.64 & 435.9 & 29.6 & 1.69 & 21.78 \\
  & {8} 
    & 1557.5 & 172.2 & 3.55 & 950.04 & 1546.3 & 216.0 & 3.51 & 1171.47 \\

\midrule

\multirow{2}{*}{Design \cite{Waris2019} } 
  &{4} 
    & 407.68 & 25.5 & 1.43 & 14.85 & 427.28 & 31.7 & 1.61 & 21.88 \\
  & {8} 
    & 1476.2 & 164.1 & 3.21 & 777.51 & 1465.2 & 207.9 & 3.18 & 966.75 \\

\midrule

\multirow{2}{*}{Design \cite{Haroon2021} } 
  &{4} 
    & 412.2 & 25.8 & 1.40 & 14.90 & 420.1 & 28.3 & 1.40 & 16.64 \\
  & {8} 
    & 1012.1 & 145.5 & 3.01 & 442.91 & 975.5 & 177.2 & 2.50 & 431.93  \\

\midrule

\multirow{2}{*}{Proposed} 
  &{4} 
    & 375.6 & 17.1 & 1.37 & 8.79 & 399.3 & 25.6 & 1.35 & 13.79 \\
  & {8} 
    & 985.2 & 125.3 & 2.71 & 334.53 & 869.5 & 155.2 & 2.48 & 334.66 \\
\bottomrule
\end{tabular}

\begin{tablenotes}
\item[\text{\makebox[0.8em][l]{$*$}}] : Reported values are normalized to 90\,nm using DeepScale tool \cite{deepscale}.

\item[\text{\makebox[0.8em][l]{$\delta$}}] : PADP unit is ($\mu\mathrm{m}^2 \!\cdot\! \mathrm{fJ}$).
\end{tablenotes}
\end{threeparttable}
\end{table}

\begin{figure}[ht]
    \centering
    \includegraphics[width=0.95\linewidth]{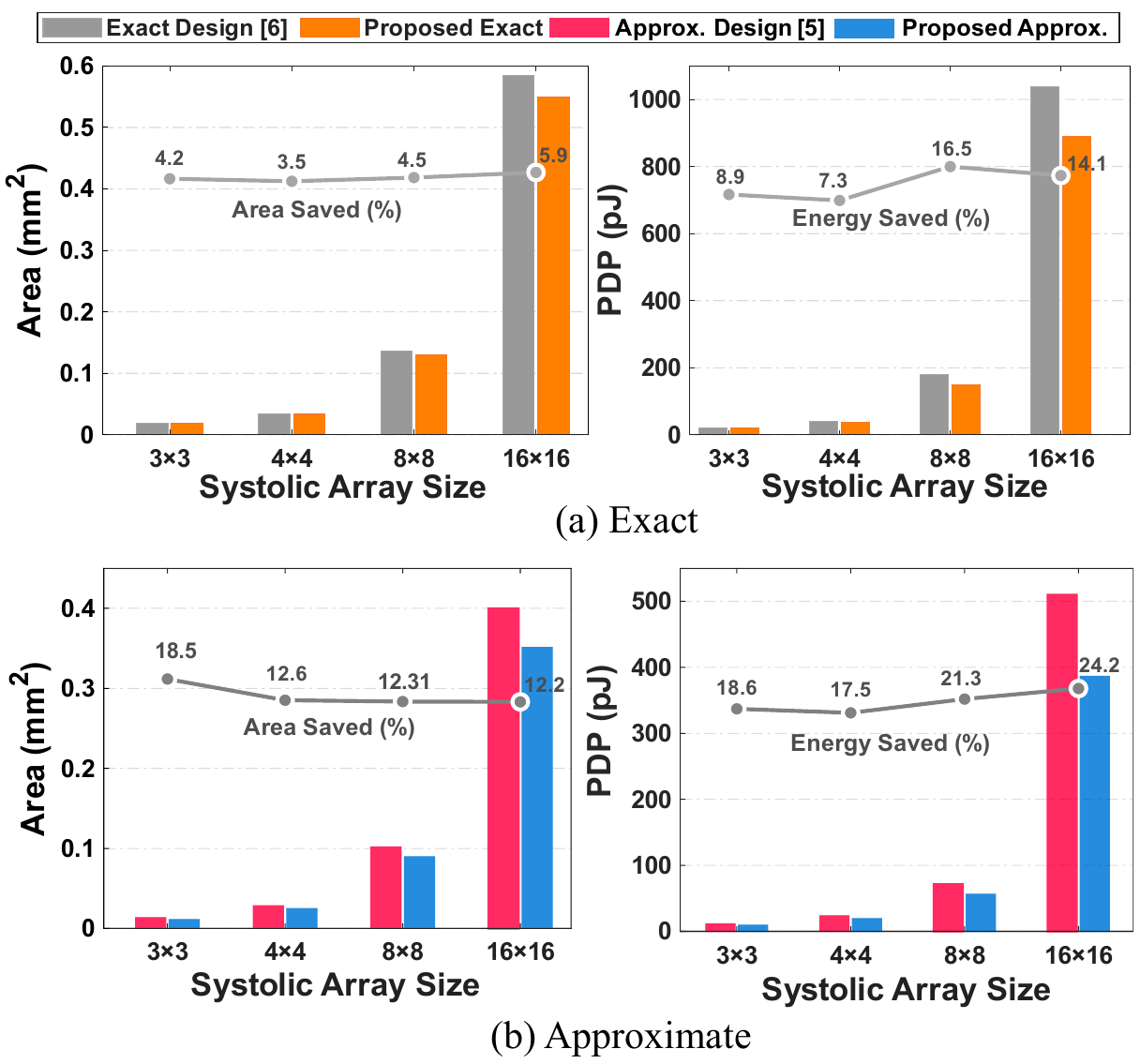}
    \caption{Area and PDP comparison of the proposed design across different systolic array sizes using 8-bit signed PE, where grey line indicating the improvement in percentage for area and energy.}
    \label{fig:PDP_SA}
\end{figure}

Table \ref{tab:sa} summarizes the results for various systolic array (SA) sizes at $250~MHz $. Fig. \ref{fig:PDP_SA} (a) shows up to 5.9\% area savings and 14.1\% energy savings of the proposed exact design over \cite{Lombardi2015} across various sizes of systolic array.
For all approximate SA configurations of varying sizes, each PE is used with an approximation factor of $k=N-1$.
For the 8-bit configuration with a 16×16 SA, the proposed approximate design achieves 
a 62.7\% reduction in PDP compared to the exact design~\cite{Lombardi2015} and a 24.2\% improvement compared to the best existing approximate design \cite{Haroon2021} as shown in  Fig. \ref{fig:PDP_SA} (b). These results indicate that the proposed approximate designs achieve substantial gains in hardware efficiency, particularly for large systolic arrays, while maintaining competitive accuracy-performance tradeoffs.

\begin{table*}
\centering
\scriptsize
\setlength{\tabcolsep}{4pt}
\caption{Hardware Analysis of Different Signed PE Designs for Various Systolic Array Sizes at 250 MHz}
\label{tab:sa}
\begin{tabular}{cccccccccccccccccc}
\toprule
\multirow{2}{*}{\textbf{bit}} & \multirow{4}{*}{\textbf{Design}} &
\multicolumn{4}{c}{\textbf{3×3}} & \multicolumn{4}{c}{\textbf{4×4}} &
\multicolumn{4}{c}{\textbf{8×8}} & \multicolumn{4}{c}{\textbf{16×16}} \\
\cmidrule(lr){3-6} \cmidrule(lr){7-10} \cmidrule(lr){11-14} \cmidrule(lr){15-18}
 \textbf{Width}& & Area & Power & Delay & PDP & Area & Power & Delay & PDP & Area & Power & Delay & PDP & Area & Power & Delay & PDP \\
 & & (mm$^2$) & (mW) & (ns) & (pJ) & (mm$^2$) & (mW) & (ns) & (pJ) & (mm$^2$) & (mW) & (ns) & (pJ) & (mm$^2$) & (mW) & (ns) & (pJ) \\
\midrule
\multirow{5}{*}{4} 
& Exact \cite{Lombardi2015}     & 0.0062 & 3.98 & 1.65 & 6.57 & 0.0112 & 3.98 & 1.67 & 6.65 & 0.0465  & 17.2 & 1.88 & 32.34 & 0.1901 & 74.4 & 2.41 & 179.30 \\
& Proposed Exact                & 0.0060 & 3.90 & 1.63 & 6.35 & 0.0110 & 3.95 & 1.64 & 5.98 & 0.0459  & 16.9 & 1.88 & 31.77 & 0.1885 & 70.7 & 2.38 & 168.26 \\
& Approx. \cite{Waris2019}       & 0.0058 & 3.89 & 1.62 & 6.30 & 0.0105 & 3.93 & 1.63 & 6.40 & 0.0445  & 16.8 & 1.87 & 31.42 & 0.1754 & 65.3 & 2.38 & 155.41 \\
& Approx. \cite{Lombardi2015}    & 0.0056 & 3.60 & 1.54 & 5.54 & 0.0101 & 3.90 & 1.50 & 5.85 & 0.0432  & 15.8 & 1.86 & 29.39 & 0.1600 & 62.80 & 2.35 & 147.58 \\
& Approx. \cite{Haroon2021}      & 0.0057 & 3.80 & 1.44 & 5.47 & 0.0103 & 3.91 & 1.30 & 5.08 & 0.0440  & 16.2 & 1.80 & 29.16 & 0.1500 & 63.00 & 2.30 & 144.90 \\
& Proposed Approx.              & 0.0050 & 3.31 & 1.40 & 4.64 & 0.0090 & 3.79 & 1.27 & 4.82 & 0.0407  & 14.3 & 1.75 & 25.19 & 0.1312 & 53.92  & 2.23 & 120.26 \\
\midrule
\multirow{5}{*}{8} 
& Exact \cite{Lombardi2015}    & 0.0191 & 6.38 & 3.36 & 21.44 & 0.0345 & 11.4 & 3.56 & 40.58 & 0.1363 & 49.8 & 3.61 & 179.78 & 0.5841 & 265.4 & 3.91 & 1037.71 \\
& Proposed Exact               & 0.0184 & 6.01 & 3.25 & 19.53 & 0.0333 & 11.0 & 3.42 & 37.62 & 0.1302 & 42.8 & 3.51 & 150.15 & 0.5498 & 233.3 & 3.82 & 891.3 \\
& Approx. \cite{Waris2019}      & 0.0155 & 5.45 & 2.97 & 16.19 & 0.0301 & 10.4 & 3.31 & 34.42 & 0.1151 & 35.1 & 3.02 & 106.00 & 0.4424 & 193.7 & 3.88 & 751.556 \\
& Approx. \cite{Lombardi2015}   & 0.0142 & 4.20 & 2.70 & 11.34 & 0.0290 & 9.60 & 2.90 & 27.84 & 0.1050 & 27.8 & 2.96 & 82.29 & 0.4200 & 166.0 & 3.70 & 614.20 \\
& Approx. \cite{Haroon2021}     & 0.0135 & 4.60 & 2.50 & 11.50 & 0.0285 & 9.20 & 2.55 & 23.46 & 0.1020 & 25.5 & 2.80 & 71.40 & 0.4000 & 150.0 & 3.40 & 510.00 \\
& Proposed Approx.             & 0.0110 & 3.86 & 2.42 & 9.36  & 0.0249 & 8.06 & 2.40 & 19.35 & 0.0895 & 20.5 & 2.74 & 56.18 & 0.3513 & 117.8 & 3.28 & 386.5 \\

\bottomrule
\end{tabular}
\end{table*}

\subsection{Error Analysis }
Accuracy of the proposed approximate MAC design is evaluated using well-known error metrics \cite{error}, Mean Relative Error Distance (MRED) and Normalized Mean Error Distance (NMED). These metrics measure the deviation of the approximate outputs from the exact. These metrics are estimated using uniformly distributed input combinations for 8-bit PE (65536) using Python simulations. Table \ref{Tab:Error} summarizes the error metrics of 4-bit and 8-bit PE designs. 
The results indicate that the accuracy of the designs gradually 
decreases with increasing approximation factor of \(k\), although 
approximations up to \(k = N\) bits still produce acceptable results as discussed in the next section.

\begin{table}[ht]
\centering
\scriptsize
\setlength{\tabcolsep}{5pt} 
\renewcommand{\arraystretch}{1.2}  
\caption{Error Metrics for Proposed and Existing 8-bit PE Designs}
\label{Tab:Error}
\begin{tabular}{cccccc}
\toprule
\multirow{3}{*}{\textbf{Design}} & \multirow{3}{*}{$\textbf{k}$ } & \multicolumn{2}{c}{\textbf{Unsigned}} & \multicolumn{2}{c}{\textbf{Signed}} \\
\cmidrule(lr){3-4} \cmidrule(lr){5-6}
  &   & NMED & MRED & NMED & MRED \\
\midrule
\multirow{4}{*}{\textbf{Proposed}} & $2$     & 0.0001 & 0.0011 & 0.0001 & 0.0037 \\
                                  & $4$     & 0.0004 & 0.0033 & 0.0004 & 0.0130 \\
\multirow{2}{*}{\textbf{Approx.}} & $5$     & 0.0006 & 0.0075 & 0.0006 & 0.0286 \\
                                  & $6$     & 0.0018 & 0.0108 & 0.0022 & 0.0481 \\
                                  & $8$     & 0.0077 & 0.0328 & 0.0081 & 0.2418 \\
    \midrule
   Design\cite{Haroon2021}        & $6$   & 0.0025 & 0.0146 & 0.0033 & 0.0626 \\
   Design \cite{Lombardi2015}     & $6$   & 0.0075 & 0.0824 & 0.0079 & 0.1064 \\
   Design\cite{Waris2019}         & $6$   & 0.0061 & 0.0250 & 0.0046 & 0.0758 \\

\bottomrule
\end{tabular}

\end{table}

\begin{figure}
    \centering
    \includegraphics[width=0.7\linewidth]{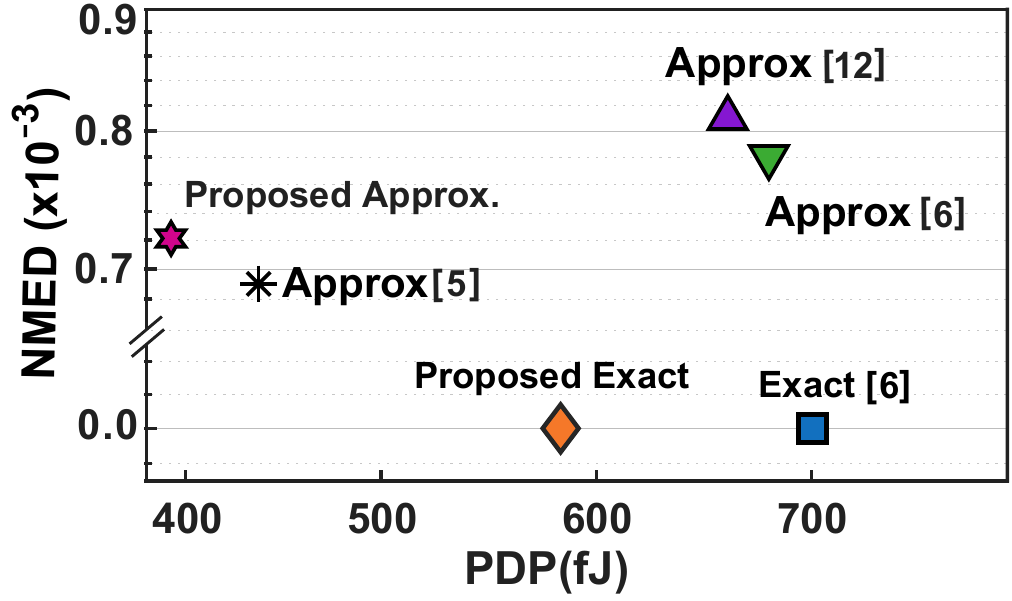}
    \caption{Comparison of PDP and NMED for Signed 8-bit PE for $k=N-1$.}
    \label{fig:scatter}
\end{figure}

\begin{figure}
    \centering
    \includegraphics[width=0.85\linewidth]{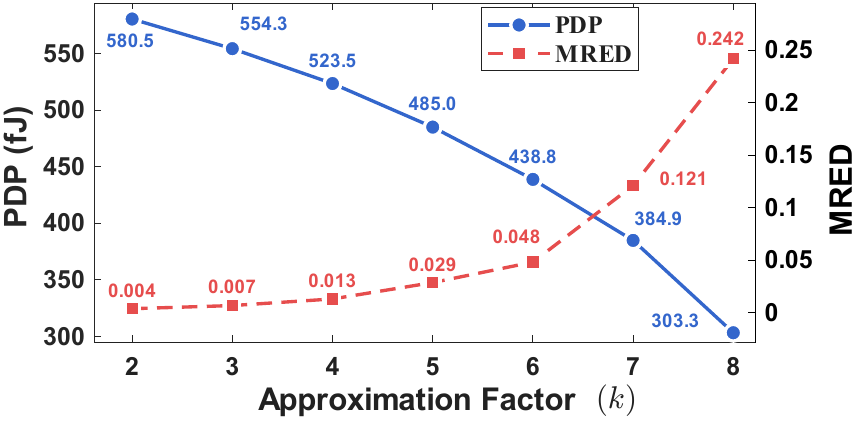}
    \caption{Variation of PDP and MRED with factor ($k$) for signed 8-bit PE.}
    \label{fig:Variation of PDP and MRED}
\end{figure}

As shown in Fig.~\ref{fig:scatter}, the proposed approximate design achieves a significant reduction in PDP while maintaining a lower NMED compared to existing approximate designs. Although the design in~\cite{Haroon2021} attains a slightly lower NMED, the proposed design offers superior area, power, and delay metrics. Furthermore, Fig. \ref{fig:Variation of PDP and MRED} presents the variation of PDP and MRED values for different approximation levels for signed 8-bit PE.

\section{Applications}
\label{sec:Application}
In this section, we evaluate the effectiveness of the proposed PEs
through three representative applications.
The first application implements an 8$\times$8 Discrete Cosine Transform (DCT) for image compression,
while the second performs edge detection using a Laplacian kernel filter. 
The third application integrates the proposed approximate PEs into a Convolutional Neural Network (CNN) based Bi-Directional Cascade Network (BDCN) architecture \cite{BDCN} for edge detection.
The image quality of the approximate PEs is evaluated using Peak Signal-to-Noise Ratio (PSNR) and Structural Similarity Index Measure (SSIM), computed with respect to the outputs of the exact design.

\subsection{Discrete Cosine Transform (DCT)}
DCT is widely used in image and video compression standards such as JPEG and MPEG. In this work, the proposed approximate PE in the systolic array performs matrix multiplication for an 8$\times$8 integer-scaled DCT. The DCT coefficient matrix is scaled to integer values \cite{DCT2014}, making it suitable for fixed-point hardware implementations, while effectively reducing computational complexity, power consumption, and area, without significantly compromising image quality.

\begin{figure}[h!]
\centering
\begin{tabular}{c c c}
\includegraphics[scale=0.21]{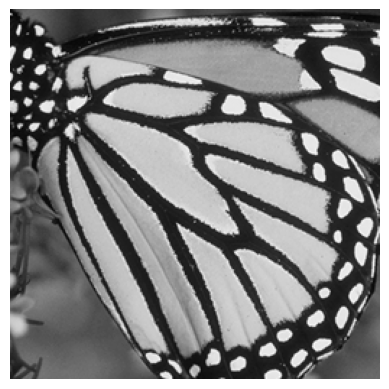} &
\includegraphics[scale=0.21]{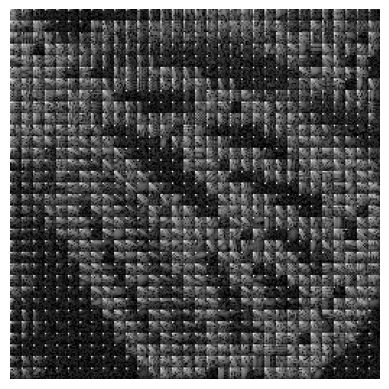} &
\includegraphics[scale=0.21]{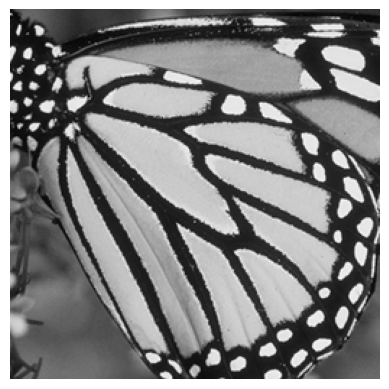} \\
\scriptsize{(a) Input image} & \scriptsize{(b) DCT coefficients} & \scriptsize{(c) Reconstructed} \\
& & \scriptsize{(45.97 dB, 0.991)} \\
\end{tabular}
\caption{Resultant images of DCT along with (PSNR and SSIM) values.}
\label{fig:dct_results}
\end{figure}

The input, DCT coefficients, and reconstructed images are shown in Fig.~\ref{fig:dct_results} along with PSNR and SSIM value. Fig.~\ref{fig:dct_results} (c) shows the output for the $k = 2$ column approximation using proposed 8-bit signed PE. As the approximation parameter $k$ increases, the output quality decreases in terms of PSNR and SSIM, as summarized in Table~\ref{tab:pe_perf}.

\begin{table}[t]
\centering
\scriptsize
\setlength{\tabcolsep}{6.5pt} 
\caption{Performance of approximate PE }
\label{tab:pe_perf}
\begin{tabular}{|c|c|cc|cc|cc|}
\hline
\textbf{PE} & \multirow{2}{*}{$\textbf{k}$} 
& \multicolumn{2}{c|}{\textbf{DCT}} 
& \multicolumn{2}{c|}{\textbf{Edge Detection}} 
& \multicolumn{2}{c|}{\textbf{BDCN-ED}} \\ \cline{3-8}
\textbf{Designs} & & PSNR & SSIM & PSNR & SSIM & PSNR & SSIM \\ \hline

Design \cite{Haroon2021}   & 8 & 27.97 & 0.854 & 10.97 & 0.538 & 28.56 & 0.865 \\ \hline
Design \cite{Lombardi2015} & 8 & 26.15 & 0.792 & 9.12 & 0.424 & 17.29 & 0.798       \\ \hline
Design \cite{Waris2019}    & 8 & 26.48 & 0.837 & 9.24 & 0.471 & 20.29 & 0.819   \\ \hline

\multirow{4}{*}{\textbf{Proposed}}  
& 2 & 45.97 & 0.991 & 30.45 & 0.910 & 75.98 & 1 \\ \cline{2-8}
& 4 & 38.21 & 0.955 & 20.51 & 0.894 & 68.55 & 1 \\ \cline{2-8}
& 6 & 35.67 & 0.923 & 12.76 & 0.678 & 51.52 & 0.999 \\ \cline{2-8}
& 8 & 28.43 & 0.872 & 11.41 & 0.651 & 34.60 & 0.995 \\ \hline
\end{tabular}
\end{table}

\subsection{Edge Detection}

\begin{figure}[h!]
\centering
\setlength{\fboxsep}{0pt} 
\setlength{\fboxrule}{0.6pt} 

\begin{tabular}{c c c c}
\hspace{-0.2cm}\fbox{\includegraphics[scale=0.2]{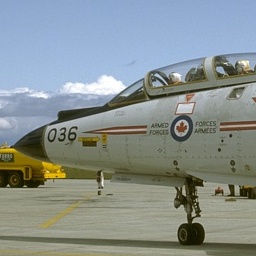}} &
\hspace{-0.3cm}\fbox{\includegraphics[scale=0.19]{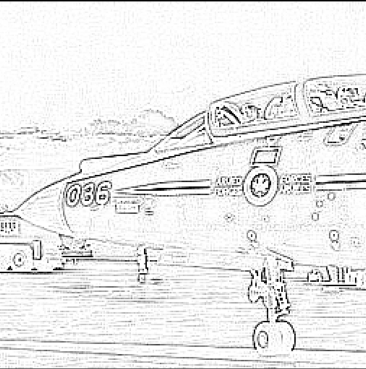}} &
\hspace{-0.3cm}\fbox{\includegraphics[scale=0.19]{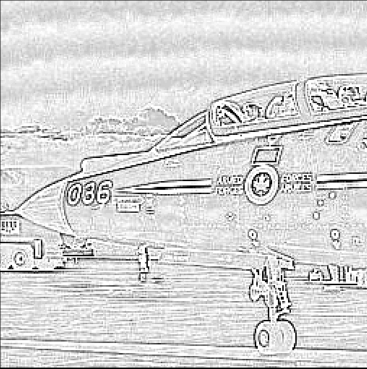}} &
\hspace{-0.3cm}\fbox{\includegraphics[scale=0.19]{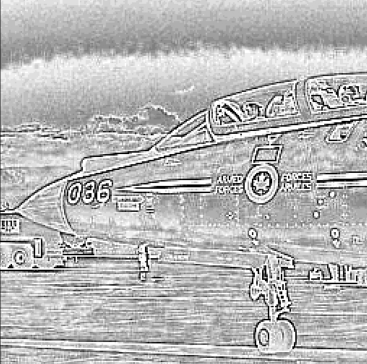}} \\
\scriptsize{Input image} & \scriptsize{Proposed exact} & \scriptsize{Proposed $k = 4$} & \scriptsize{Proposed $k = 6$} \\
& \scriptsize{using kernel} & \scriptsize{PSNR: 20.51dB} & \scriptsize{PSNR: 12.76 dB} \\[4pt]

\hspace{-0.2cm}\fbox{\includegraphics[scale=0.2]{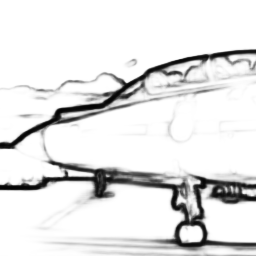}} &
\hspace{-0.3cm}\fbox{\includegraphics[scale=0.2]{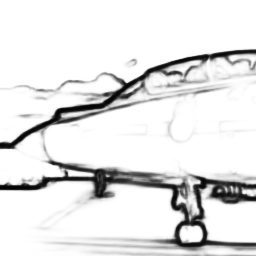}} &
\hspace{-0.3cm}\fbox{\includegraphics[scale=0.2]{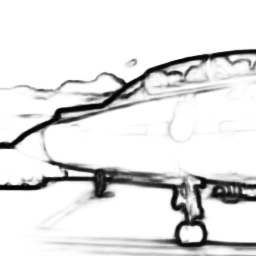}} &
\hspace{-0.3cm}\fbox{\includegraphics[scale=0.2]{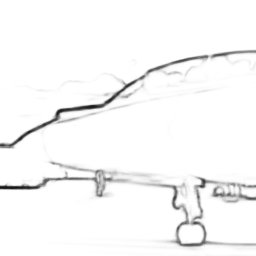}} \\
\scriptsize{Proposed exact} & \scriptsize{Proposed $k = 2$} & \scriptsize{Proposed $k = 8$} & \scriptsize{Design \cite{Waris2019}} \\
\scriptsize{using BDCN} & \scriptsize{PSNR: 75.98 dB} & \scriptsize{PSNR: 34.60 dB} & \scriptsize{PSNR: 20.29 dB} \\
\end{tabular}

\caption{Comparison of edge detection results, where the first row presents the outputs generated by the Laplacian kernel, while the second row illustrates the corresponding results produced by the BDCN model.}
\label{fig:edge_results}
\end{figure}

The proposed approximate signed PE is evaluated for edge detection using both conventional laplacian kernel and a CNN-based BDCN \cite{BDCN}. In the kernel-based approach, the laplacian filter is convolved using different approximation levels. Lower $k$ values provide high PSNR and SSIM with good visual quality, while higher $k$ values yield major degradation in the image quality as shown in Fig. \ref{fig:edge_results} first row.

\begin{figure}[h!]
    \centering
    \includegraphics[width=1\linewidth]{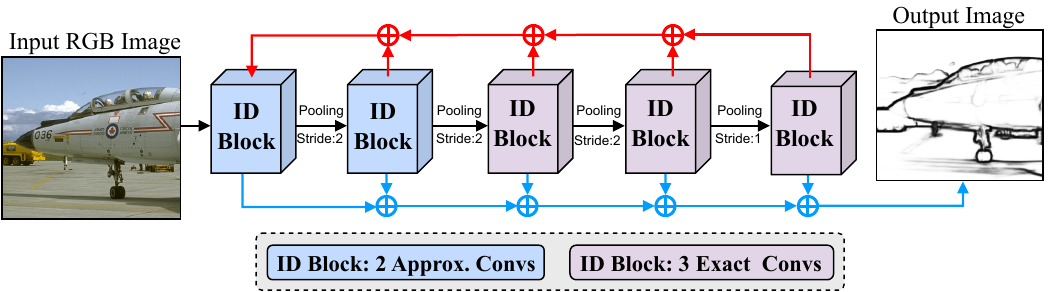}
    \vspace{-5mm}
    \caption{Approximate Architecture of BDCN \cite{BDCN}.}
    \vspace{-3mm}
    \label{fig:archi_BDCN}
\end{figure}

In the CNN-based approach, the proposed architecture is a modified version of the conventional Bi-Directional Cascade Network for perceptual edge detection (BDCN) model \cite{BDCN}
as shown in Fig. \ref{fig:archi_BDCN}. 
This network consists of a sequence of convolution, 
pooling, and upsampling layers arranged in a bi-directional cascade form. 
The first two blocks  employ approximate PEs 
to reduce computation cost and power consumption, 
whereas the subsequent blocks maintain 
full-precision computations to preserve accuracy in the later feature extraction stages.
This hybrid design effectively balances accuracy and efficiency, as the early layers can tolerate small computation errors. 

As observed from Table \ref{tab:pe_perf} and Fig. \ref{fig:edge_results}, the CNN-based edge detection  consistently achieves higher PSNR and SSIM than the kernel-based method. Even for higher approximation factors ($k=6,8$), BDCN maintains superior image quality, demonstrating that the network structure efficiently compensates for arithmetic inaccuracies while achieving significant power and area savings. This validates the effectiveness of integrating approximate PEs within CNN-based edge detection frameworks, highlighting their potential for energy-efficient deep learning hardware implementations.

\section{Conclusion}
\label{sec:conclusion}
This work presents an energy efficient systolic array architecture leveraging proposed exact and approximate PE with optimized PPC and NPPC blocks. The proposed designs demonstrate substantial energy savings of up to 68\% while maintaining high output quality, validated through applications such as, DCT based image compression and convolution based edge detection. These results highlight the suitability of the architecture for error-resilient image and vision processing tasks, offering a promising balance between performance, energy efficiency, and computational accuracy.

\bibliographystyle{IEEEtran}
\bibliography{IEEEabrv,main}

\end{document}